\documentclass[conference,a4paper]{IEEEtran}

\newcommand{\remove}[1]{}
\newtheorem{definition}{Definition}
\newtheorem{lemma}{Lemma}
\newtheorem{theorem}{Theorem}
\newtheorem{corollary}{Corollary}
\usepackage[cmex10]{amsmath}
\usepackage{amssymb}
\usepackage{mathrsfs}
\usepackage{graphicx}
\usepackage{color}
\usepackage{multirow}
\usepackage{amssymb}
\usepackage[top=19mm, bottom=43mm, left=13mm, right=13mm]{geometry}
\usepackage{framed}

\newcommand{\calM}{{\cal M}}

\begin{document}

\sloppy

\title{Efficient Codes for Limited View Adversarial Channels }

\author{
  \IEEEauthorblockN{Reihaneh Safavi-Naini, Pengwei Wang}
 \IEEEauthorblockA{Department of Computer Science\\
    University of Calgary\\
    Calgary, Canada\\
    Email: [rei, pengwwan]@ucalgary.ca} 
   

}



\maketitle

\begin{abstract}
We introduce randomized Limited View (LV) adversary codes that provide protection against an
adversary that  uses their partial view of the communication to construct an adversarial
error vector  to be added to the channel.  
 For a codeword of length $N$, the adversary  
selects a subset of
$\rho_rN$  of the codeword components to ``see", and then ``adds"  an adversarial error vector of
weight $\rho_wN$ to the codeword.
Performance of the code is measured by the probability of the decoder failure in recovering the sent message. 
 An $(N, q^{RN},\delta)$-limited 
view adversary code ensures that the success chance of the adversary in making decoder fail, is bounded by $\delta$ when the  information rate of the code is at least $R$.
Our main motivation to study these codes is providing protection for wireless communication at the
physical layer of networks.

We formalize the definition of adversarial error and decoder failure, construct a code   with efficient encoding and decoding that allows the
adversary to, depending on the code rate, read up to half of the sent codeword and add error on the same  coordinates.
The code is non-linear,   has an efficient decoding algorithm, and is constructed using a message authentication code (MAC) and a
Folded Reed-Solomon (FRS) code. 
The decoding algorithm 
uses an innovative approach that combines the list decoding algorithm of the  FRS codes and the MAC verification
algorithm to eliminate the exponential size of the list output from the decoding algorithm.
We discuss application of our results to Reliable Message Transmission problem, and open problems for future work.

\end{abstract}

\section{Introduction} \label{intro}
Shannon \cite{S48}  formalized the study of reliable communication over noisy channels where  transmitted symbols
are changed  according to a known fixed probability distribution.  In adversarial channels corruption of transmitted
symbols is adversarial: the adversary can corrupt any subset of the symbols as long as the size of the set is bounded
and is a constant fraction of the transmitted sequence.
Much less is known about adversarial channels.  For example, although  it is well known that the information
 capacity of a binary symmetric channel 
with crossover probability $\rho$ is $1-H(\rho)$, the answer to the same question in the case of binary adversarial channels where the adversary  corrupts a $\rho$ fraction of bits in unknown, although it is known that it is much less than $1-H(\rho)$.
Adversarial channels have received much attention in recent years  \cite{GS10}\cite{M04}\cite{M08}
as they provide a powerful method of modelling  communication channels where 
the channel behaviour is not known or varies over time.

In adversarial channels, one commonly assumes that the sent codeword is 
known, or even chosen (for example in randomized codes) by the adversary and that the adversary is allowed to corrupt a fraction of the sent symbols. For {\em unique decoding} the number of errors must be less than half the minimum distance of
the code, and 
 for  higher fraction of errors, one needs to make extra assumptions such as a 
 secret key shared by the sender and receiver in {\em private codes} \cite{M04}, or bound on the computation of the adversary  \cite{MPSW05}.

In this paper we consider an adversary with unlimited computation but  assume that the adversary has a {\em limited view} of the transmitted codeword.
That is we assume the adversary can see only a fraction  of the sent codeword and can add errors to a fraction, possibly different,  of the codeword.
In other words the adversarial capability is specified by a pair of parameters $(\rho_r, \rho_w)$,
meaning that the adversary can read $\rho_r N$ components of their choice, and corrupt
$ \rho_w N$ components of their choice. We do not assume any shared secret key.

\subsection{Motivations} 
One of the motivations of our work is  to model an on-line adversary in a wireless communication system,
where the adversary can partially observe the communicated symbols before tampering with them \cite{PTDC11}.

We assume the encoded message is a $q$-ary vector and that the adversary can choose the positions that
he would like to ``see" (the remaining positions are not visible to the adversary) and then designs the
tampering vector (noise) that is ``added" to the encoded message.
Our definition of limited view adversary codes aims to {\em guarantee reliable authentic communication  at the physical layer of communication channels}
and this means  that the decoder will never output an incorrect (un-authentic) message,  and with a very small probability fails to output the correct message.
A somewhat similar scenario is 
has been considered in Algebraic Manipulation Detection Codes (AMD)  \cite{CDFP08} where the encoded message is
stored in a secure storage and the adversary can only ``add" errors 
to the codeword.  In AMD codes the adversary cannot ``see" the stored 
codeword and the aim of the code is to {\em detect} tampering with the message.
We allow some partial information to be ``leaked" to the adversary and
the goal of the coding is to correctly recover the message.
Note that because the code is randomized, recovering the message does not imply that the added noise can be found.

A second motivation for our model is to study  {\em 1-round $\delta$-Reliable Message Transmission (RMT)} \cite{DDWY93} as a code and so establish the relationship between two seemingly different areas of communication over networks, and communication over noisy channels. Such relationship can enrich the tools and techniques developed in each area
and   result in better understanding and constructions in the two cases.
In  RMT scenario a sender is connected to a receiver through a set of $N$ node disjoint communication paths,
a subset of which is controlled by an adversary who can see what is  sent  on a controlled
path and can replace it with a value of their choosing.  Communication paths in RMT scenario are assumed end to end and {\em unlike network coding} \cite{ACLY00}, nodes
in the network do not take part in the communication protocol.
In RMT the information processing is by the legitimate users (encoding and decoding)  and happens   at the ends of a
 path.  The adversary interacts with the system  by reading a subset of paths and changing the 
value sent over another subset  of paths. {\em When the two subsets are the same, the modification can be represented as
adding an error vector.} 
  $\delta$-RMT protocols  in general are multi-round and guarantee that message is correctly received with a probability 
at least $1-\delta$. The bulk of research on $\delta$-RMT protocol assumes the adversary 
reads and modifies the same subset of paths.

\subsection{Our work}
We define and formalize randomized (stochastic) limited view adversary codes, with security against 
 an adversary  who can choose a fraction of positions of codeword to read and then add errors.   
For codewords of length $N$, a $(\rho_r, \rho_w)$ adversary  
selects a subset of
$\rho_rN$ components to see, and then adds (component-wise addition over $F_q$) an error vector of
weight $\rho_wN$ to the codeword.
The decoder outputs either the correct message or a symbol $\perp$, that shows the decoder failure.
Performance of a code is measured by the probability of the decoder outputting $\perp$; this is the
 success probability of the adversary in making the decoder fail. An $(N, M, \delta)$-LV adversary code
guarantees that the message can be correctly recovered against a  $(\rho_r, \rho_w)$ adversary, and the
success chance of the adversary in making the decoder to fail is upper-bounded by $\delta$.
The information rate of a code of length $N$ with $M$ codewords is $\frac{\log_q M}{N}$.
A good code will have high information rate for  high values of $\rho_r$ and $ \rho_w$.

We construct an $(N, M, \delta)$-LV adversary code that is non-linear, and uses two building blocks: a message authentication code and a
Folded Reed-Solomon (FRS) code. To encode a message $m$, the sender first chooses $N$ appropriately constructed secret keys,  uses the keys to construct $N$ authentication tags for the message using the chosen MAC
 (See MAC Construction II for details), and appends the tags to the message. 
The tagged message is then encoded using an FRS code. The $i^{th}$ component of the final codeword which is sent to the receiver consists of the corresponding component of the FRS code and the MAC key.
The decoder recovers the correct message in a conceptually two step process:
using the list decoding algorithm of the FRS code to construct a list of possible codewords and 
then applying the MAC verification algorithm to output either the correct message, or $\perp$.
 This two step algorithm however can result  in an exponential cost decoding because the  output list of the FRS decoding algorithm  
can be  of exponential size. A previous application of the general approach  of using MACs and
FRS codes for the construction of 1-round RMT \cite{STW12} has this shortcoming.
The innovation in this paper is to combine the system of linear equations resulting from
the algebraic list decoding algorithm \cite{Gur11} of FRS codes, with a set of linear equations resulting
from the verification algorithm of a specially constructed MAC, to have a 
single system of linear equation whose solution gives the correct message with a high probability. 
The MAC in this construction must be a key efficient MAC that 
can be used for different length messages and have appropriate verification algorithm suitable for efficient decoding. 
MAC Construction II satisfies these properties and could be of
independent interest. 
{\em The final decoder  complexity is polynomial.}

The code allows the
adversary to, depending on the code rate, read up to half of the codeword and adds error  on the same number of coordinates. 

\smallskip
\noindent
{\em RMT Construction:}
One of the motivations for defining LV adversary codes is to cast the 1-round $\delta$-RMT construction as a 
coding problem. Our construction of 
LV adversary code   can  be immediately used to give an optimal 1-round $\delta$-RMT 
construction (See Section \ref{RMT} for definitions.) whose parameters match the best known RMT constructions  \cite{STW12}.
It is interesting to note that the LV adversary code parameters provide a more refined set of parameters for the evaluation
of RMT. In particular,   
a 1-round $\delta$-RMT is optimal if
 transmission rate is $O(1)$.  Noting that transmission rate in RMT is the inverse of the information rate (See Section \ref{RMT})
in LV adversary codes, any LV adversary code with non-zero information rate immediately results in an optimal 1-round $\delta$-RMT. 
For LV adversary codes however 
the rate of information communication is a key efficiency parameter and the goal is to maximize this 
rate (with other parameters fixed).  LV adversary code view of  1-round $\delta$-RMT  allows
 comparison of optimal systems in terms of their information rate.
    In addition to providing efficient decoding, the LV adversary code construction in this paper
allows the parameters of the 1-round $\delta$-RMT code to be chosen such that the protocol achieves maximum information rate.

LV adversarial channels and codes open many new open questions.  Finding general bounds and relationship
among the information rate $R$, observation and corruption ratios,  $\rho_r$ and $\rho_w$ respectively, and
finding the highest information rate (capacity) of LV adversary codes remain important research questions.
Also construction of good codes by refining our approach here (combining message authentications codes and  list decodable codes),
or using new  approaches, are interesting open problems.

\subsection{Related work}

In a previous  submission \cite{SW13}  we introduced deterministic LV adversary codes and gave a deterministic construction of 
such codes. Deterministic  encoding  enforces restrictions on 
$\rho_r$ and $\rho_w$, that can be overcome  by the randomized codes.  
The definition of decoder error in this paper follows the same approach as deterministic
codes, but is in terms of probabilities instead of the combinatorics of the code.
This is needed because of the randomize nature of the code removes restrictions that are dictated
by the deterministic  (one message, one codeword) nature of the code.
In the same submission we also showed how to adapt a 1-round RMT protocol in \cite{STW12} to construct a
  randomized construction for limited view codes.
Decoding complexity of this construction was exponential and no security model and proof was provided for 
the code.

Protection against message manipulation was first considered in \cite{CW77} and later formalized 
as message authentication codes 
in \cite{S85}. As noted earlier message authentication codes require shared secret key and
provide protection against a powerful  adversary who can completely replace a sent coded message with
another one.  The  security guarantee for these codes is {\em detection} of manipulation.

Adversarial tampering by an adversary that does not  ``see" the encoded message, has been considered in \cite{CDFP08}.
AMD codes  do not need a secret key but tampering is only by adding an adversarial noise.
LV adversary codes do not require shared secret and aim at recovering the message.
They limit manipulation to adding the nose but  allow adversary to
partially see the codeword 
before designing their adversarial noise vector.

Adversarial channels have been widely studied in the literature \cite{CN88}, \cite{LN98}. Our model of adversarial channel has similarity 
with the model in \cite{M08} where {\em binary} oblivious channels are introduced. 
In  oblivious channels the adversary sees the codeword, and depending on the
level of obliviousness, can use one of the limited number of distributions on the error vectors that are available to them.
A  $\gamma$-oblivious adversary 
can emply at most $2^{1-\gamma}$ error distributions for corrupting the codewords.
In these codes  each codeword is associated with one error distributions. 
By limiting the adversary's reading capability, our limited view adversary also effectively limits  
the number of distributions that the 
adversary can use. However each codeword can have more than one error distributions.

{\it Organization.} 

In Section 2, we give the background for Folded Reed-Solomon code,  1-round $\delta$-RMT codes and message authentication codes. 
In  Section 3, we introduce the randomized limited view adversary code and give new constructions for MAC. In Section 4, we present an efficient construction for randomized limited view adversary code. Section 5 discusses our results,   open problems and future works.

\section{Background}
We give an overview of the main building blocks and definitions required in this paper.

\subsection{Folded Reed-Solomon code} \label{FRS_Code}

Error correcting codes are used for reliable data transmission over noisy channels.
Let the message space be a set ${\cal M} $ with  probability distribution $\Pr(m)$.

\begin{definition}
An $[N, q^{RN}]$ error correcting  code $C$ with information rate $R$, is a set of $q^{RN}$ code vectors $C=\{c_1, \cdots, c_{q^{RN}}\}$ where  $c_i\in F_q^N$.
The code has two algorithms:
an encoding and  a decoding algorithm. 
The encoding algorithm $Enc: {\cal M}  \rightarrow C$  maps a message from $\cal M$ to a codeword in $C$ that is sent over the channel.
 The decoding algorithm  $Dec: F_q^N \rightarrow  {\cal M}\cup \{\perp\}$  is a deterministic algorithm that takes any vector in $F_q^N$ and outputs a message in $ {\cal M}$ or fails, outputting a symbol $\perp$.
 A decoder error occurs if $Dec(Enc(m,r))\neq m$.
 \end{definition}

The Hamming weight of a vector $e\in F_q^N$ is denoted by $wt(e)$ and is the number of non-zero components of $e$. For a vector $y\in F_q^N$ and an integer $r$, let $B(y, r)$ be the Hamming ball of radius $r$ centred at $y$. 
Let $\rho$ denote the fraction of errors (the number of errors divided by the length of the codeword)  that can be corrected by the decoder.

\begin{definition}
A {\em Bounded Distance Decoding (BDD)} algorithm $Dec(y)$ takes a received word $y=(y_1, \cdots, y_N)$ and outputs $m\in {\cal M}$ 
if $m$ is the unique message of the codeword(s) that are at distance at most $wt(e)$ from $y$.
 The decoder outputs $\perp$  otherwise.

\remove{
\[
Dec(y)= \left\{\begin{matrix}
\begin{split}
& m\ \ \ & if\ \{ |m: B(y, wt(e))|=1\\
&\perp & if\ |m: B(y, wt(e))|>1
\end{split}
\end{matrix}\right. 
\]
}
For deterministic codes, the above definition implies that the decoder outputs $m$, if $Enc(m)$ is the only codeword in $ B(y, wt(e))$. In randomized codes however, $ B(y, wt(e))$ may contain more than one encoding of $m$.

Using  bounded distance decoding, 
the receiver $\cal R$ outputs either
 a message $m$ or the fail symbol $\perp$, that is $Dec(y) \in \{ \calM, \perp\}$.
\end{definition}

The above decoding is a {\em  unique decoding} algorithm
and requires that the  output  is a single  message, or the fail symbol. 
For this decoding,  correct decoding can be guaranteed if
$\rho$ is less than half of the minimum distance of the code, 
that is $\rho\leq \frac{1-R}{2}$. Reed-Solomon code has an efficient unique decoding algorithm that
can correct at most a fraction $\rho = \frac{1-R}{2}$ errors.

\begin{definition}
An $(N, k)$ Reed-Solomon code with block length $N (< q)$ and dimension $k$ over field  $F_q$, is a linear code with encoding and decoding described below.  
A message block of length $k$ defines  a  polynomial $f(x)$ of degree at most $k-1$ over $F_q$.
The codeword corresponding to this message block is the vector obtained by the
evaluation of this 
 polynomial  
 at $N$ distinct values $\alpha_1,\cdots, \alpha_N$, where $\alpha_i\in F_q, i=1\cdots N$. 
 That is the  codeword is $(f(\alpha_1),\cdots, f(\alpha_N))$.
\end{definition} 

For higher error ratios, one can use {\em list decoding} \cite{Eli57}
where the decoder outputs a list of possible codewords (messages).

\begin{definition} \label{LisDec}
Let $(N, q^{RN})$ code to be a code with length $N$ and information rate $R$. A code $C$  is $(\rho,L)$-list decodable if the number of codewords within distance  $\rho N$ of any received word is at most $L$. That is for every word $y \in q^N$, there are at most $L$ codewords at distance $\rho N$ or less from $y$.
List decodable codes can potentially correct up to 
$1-R$ fraction of errors. This is twice 
that of unique decoding and is called the {\em list decoding capacity} of the code.
\end{definition}

Construction of  good codes with efficient list decoding algorithms is an important research question.
An explicit construction of list decodable code that achieves the list decoding capacity $\rho=1-R-\varepsilon$ is given by Guruswami et al. \cite{Gur11}. The code is called {\em Folded Reed-Solomon codes (FRS codes)} and has polynomial time encoding and decoding algorithms.

\begin{definition} 
A $u_1$-folded Reed-Solomon code is a code with block length $N=n/{u_1}$ over $F_q^{u_1}$ with $|F_q|>n$. We represent the message by a polynomial $f(x)$ of degree at most $k$ over $F_q$,
The FRS codeword is  over $F_q^{u_1}$ and each  of its component is a $u_1$-tuple $(f(\gamma^{ju_1}), f(\gamma^{ju_1+1}), \cdots, f(\gamma^{ju_1+u_1-1}))$,  for $0\leq j<N$, where $\gamma$ is a generator of $F_q^*$.
In other words  a codeword of a $u_1$-folded Reed Solomon code of length $N$ is  in one-to-one correspondence with  a codeword $c$ of a Reed Solomon code of length $u_1N$, and is obtained by grouping together $u_1$consecutive components  of $c$. 
\begin{equation}\label{FRS_encode}
\begin{bmatrix}
f(1) & f(\gamma^{u_1}) & \cdots & f(\gamma^{u_1(N-1)})\\
f(\gamma) & f(\gamma^{u_1+1}) & \cdots & f(\gamma^{u_1(N-1)+1})\\
\vdots & \vdots & \ddots  & \vdots \\
f(\gamma^{u_1-1}) & f(\gamma^{2u_1-1}) & \cdots & f(\gamma^{u_1N-1})
\end{bmatrix}
\end{equation}
\end{definition}
We denote the encoding algorithm of FRS code by $Enc_{FRS}$. $u_1$ is called the {\em  folding parameter} of the FRS code. 

There are a number of  efficient  list decoding algorithms for FRS codes. We will use the {\em linear algebraic FRS decoding algorithm} \cite{Gur11}. The algorithm
 reduces the list decoding problem of the code to solving a set of linear equations. 
 This algorithm, although not the best in terms of the number of corrected errors, but asymptotically achieves the list decoding capacity. The structure of the decoding algorithm of the FRS code makes it possible to combine it with the new MAC verification algorithm, to obtain  an efficient decoding algorithm for the LV adversary code. 
 The following Theorem gives the decoding capability of linear algebraic FRS code.

\begin{lemma} \cite{Gur11} \label{le_fd} For the Folded Reed-Solomon code of block length $N $ and rate $R = \frac{k}{u_1N}$, the following holds for all integers $1\leq v\leq u_1$. Given a received word $y \in (F_q^{u_1} )^N$, in $O((Nu_1\log q)^2)$ time, one can find a basis for a subspace of dimension at most $v - 1$ that contains all message polynomials $f \in F_q[X]$ of degree less than $k$ whose FRS encoding agree with $y$ in at least a fraction,
\[
N-\rho N>N(\frac{1}{v+1}+\frac{v}{v+1}\frac{u_1R}{u_1-v+1})
\]
of $N$ codeword positions. The algorithm outputs a list of size at most $q^{v-1}$.
\end{lemma}
The decoding algorithm of FRS code is in appendix \ref{decode_FRS}.

\subsection{Reliable Message Transmission} \label{RMT}

In a 1-round $\delta$-RMT problem, the sender $\cal S$ and the receiver $\cal R$ are connected by $N$ node disjoint paths.
The goal is to enable $\cal S$ to send a message $m$, drawn from message space $\cal M$ to $\cal R$ such that $\cal R$ receives the message \emph{reliably}.
The adversary $\cal A$ has unlimited computational power and in threshold RMT, can corrupt any subset of 
 at most $t$ out of the $N$ paths which is unknown to $\cal S$ and $\cal R$: the adversary can eavesdrop, block or modify communication that is sent over the corrupted wires. 
$\cal S$ uses the {\it encoding algorithm} of the RMT protocol to encode the message $m$ into transcript 
that is sent to $\cal R$. 
The transcript  may be corrupted by $\cal A$ and is received by $\cal R$ who uses the {\it decoding algorithm} of the RMT 
protocol to output a message $m$, or output $\perp$.

\begin{definition}\label{def_rmt}
An RMT 
protocol between $\cal S$ and $\cal R$ is 1-round $\delta$-reliable message transmission ($\delta$-RMT) protocol if $\cal{R}$ correctly receives the message $m$ with probability $\ge 1- \delta$, and outputs $\perp$ with probability $\leq \delta$. 
The receiver never outputs an incorrect message:
\[
\Pr[ {\cal R}\ outputs \perp ] \leq \delta
\]
\end{definition}

The transmission efficiency is measured by the {\em transmission rate}
which 
is the ratio of the total  number of bits 
 transmitted from $\cal S$ to $\cal R$ to the length of the message in bits. 
Protocols whose {\it transmission rate} asymptotically matches the lower bounds are called \textit{ optimal.} Optimal 1-round $\delta-$RMT protocols must have transmission rates $\mathcal{O}(1)$.

{\it Computational efficiency} is measured by the computational complexity of the encoding and the decoding,
as a function of $N$. Efficient scheme needs polynomial (in $N$) computation of both encoding and decoding algorithm. 

\subsection{Message authentication codes}
A message authentication code (MAC) is a  cryptographic primitive that allows a sender who shares a secret key with 
the receiver to send an information block over a channel that is tampered by an adversary, enabling
the receiver 
to  verify the integrity of the received message. 
We follow the terminology of \cite{S85} and refer to the information block as {\em source state}, and to the 
authenticated message  that is sent over the channel as, the {\em message}.
A message authentication code  consists of
two algorithms $(MAC;Ver)$ that are used for tag generation and verification, respectively. 
 The sender of a source state $x$ computes an {\em  authentication tag,}  or simply a {\em tag}, $y = MAC(k;x)$, and forms the message $(x,y)$ to be sent over the channel.  
The receiver accepts the pair $(x,y)$ if $Ver((x, y), k)) = 1$.
Security of a 1-time MAC is  by requiring,
$$\Pr [(x', y'), Ver(k, (x',y') ) =1|(x, y),  y= MAC(k, x) ]\leq \varepsilon$$

\section{Model, Definitions and Building Blocks}


We first introduce our model of randomized LV adversarial channel, and define the decoding error for randomized LV adversary codes. We then  describe  the construction of a new message authentication code with provable security, that is used in the construction of the LV adversary code.

\subsection{Limited view adversary}

An $(N, M)$ randomized LV adversary code  $C$  of length $N$ with $M$ codewords over $F_q$,
consist of a probabilistic encoding algorithm, $Enc: {\cal M}\times U \rightarrow C$, from a
message set $\cal M$ of size $M$ to a code book $C$.  Here $U$ is the randomness used in the encoding.
The encoding and decoding algorithms are $Enc(m,r)$ and
$Dec(y) \in\{ {\cal M} \cup \perp\}$, respectively. Let   $C^{m}=\{c: c=Enc(m, r), \forall r\in U\}$. 
To guarantee perfect decodability without error, we assume  $C^{m} \cap C^{m'} =\emptyset,\,\, m\neq m'$.

Let $[N]=\{1,\cdots, N\}$, and  $S_r= \{i_1,\cdots, i_{\rho_rN}\} \subset [N]$  and   $S_w= \{j_1,\cdots, j_{\rho_wN}\} \subset [N]$ be two subsets
of positions.

\begin{definition}
A $(\rho_r, \rho_w)$ limited view adversary, or a $(\rho_r, \rho_w)$ LV adversary for short,  has two capabilities: reading and writing.
For a codeword of length $N$, these capabilities are: 
\begin{itemize}
\item Reading: Adversary reads a subset $S_r$ of size $\rho_rN$, of the components of the sent codeword $c$ and learns, $(c_{i_1},\cdots, c_{i_{\rho_rN}})$.
\item Writing: Adversary  adds (component wise and over $F_q$) to the sent codeword, an error vector $e$ with $wt(e)= \rho_wN$, whose non-zero components are on $S_w$. The corrupted components of $c$ in $S_w$ are,
$(y_{j_1},\cdots, y_{j_{\rho_wN}})$.
\end{itemize}

\end{definition}

The adversary is {\em adaptive}: 
 that is the adversary first chooses $i_1$ to see, and based on the seen
value $c_{i_1}$, chooses $i_2$ and so on. That is to choose any member of $S_r$, the adversary uses the knowledge of all the components that have been seen till then. The adversary then adaptively chooses $S_w$, and the error vector $e$.

\subsection{Randomized limited view adversary code}

By observing the values $\{c_{i_1},\cdots, c_{i_{\rho_rN}}\}$,
the adversary can  determine  a subset of possible sent codewords (those that match the seen positions).
Let  ${\cal C}[c_{i_1},\cdots, c_{i_{\rho_rN}}]$ denote the set of codewords that have $\{c_{i_1},\cdots, c_{i_{\rho_rN}}\}$ in positions $S_r= \{i_1,\cdots, i_{\rho_rN}\}$.

\subsubsection{Decoding error}

Decoder uses bounded distance decoding with radius $\rho_wN$: for a received vector $y$, it considers all 
codewords that are in $ B(y, \rho_wN)$ and if it finds encodings of a {\em unique} message, it outputs that message; Otherwise it outputs $\perp$.
The error vector $e$ is of weight  $w_H(e)\leq \rho_wN$
and is chosen by the adversary after reading $\{c_{i_1},\cdots, c_{i_{\rho_rN}}\}$. 
The adversary can find the failure probability of the decoder for any error vector $e$,   and choose the ``best" one;  this is the $e$ that results in the highest failure  probability for the decoder.

\begin{definition}\label{def_generaldecodingerror2}
Consider an additive error $e$ with $w_H(e)= \rho_wN$.
The decoding error $\delta_e({\cal C}[c_{i_1},\cdots, c_{i_{\rho_rN}}])$  
for a message $m$ and an error $e$ if adversary chooses to read a $S_r$ and see $\{c_{i_1},\cdots, c_{i_{\rho_rN}}\}$ in those positions is
\[
\begin{split}
&\delta_e({\cal C}[c_{i_1},\cdots, c_{i_{\rho_rN}}])=\Pr[ Enc(m, r)\in {\cal C}[c_{i_1}\cdots c_{i_{\rho_rN}}] \\
&\wedge Dec(Enc(m,r)+e)=\perp |\ {\cal C}[c_{i_1}\cdots c_{i_{\rho_rN}}]   ]
\end{split}
\]
The decoding algorithm fails, that is $Dec(Enc(m,r)+e)=\perp$, if and only if there exist $c'\in C\setminus {C}^{m}$ and $c'\in B(c+e, \rho_rN)$.

The decoding error for the decoder is,
\[
\begin{split}
\delta= \max_{S_r}\max_{c_{i_1},\cdots, c_{i_{\rho_rN}}} \max_e \delta_e({\cal C}[c_{i_1},\cdots, c_{i_{\rho_rN}}])
\end{split}
\]
\end{definition}

\begin{definition}
An $(N, M, \delta)$ randomized LV adversary code with protection against $(\rho_r, \rho_w)$ adversary, ensures that the probability of the decoding failure defined as above, is no more than $\delta$.
\end{definition}

\subsection{MAC Construction }

In the following we first give Construction I for a MAC, and then in Section \ref{MAC II} give Construction II which is an
equivalent polynomial representation for it.  This latter MAC will be used in the construction of the LV adversary code in Section \ref{SMT2}. Construction I provides an intuitive understanding of 
Construction II. 

Both MACs are $\frac{2}{q^N}$ secure. 

\subsubsection{MAC Construction I} 

The MAC is defined over $F_{q^N}$ and works for any length message.
The source state of the MAC is ${\bf x}=(x_1,\cdots, x_{l})$, where $l$ is  any integer and $l>0$. 
The MAC key  is ${\bf r}=(r_1,\cdots, r_{d}, r_{d+1})$ where $d$ is the smallest integer that satisfies $\frac{d(d+3)}{2}\geq l$. The message of MAC is $({\bf x}, tag)$. The tag generation is given by,
\[
\begin{split}
tag=&MAC({\bf x}, {\bf r})=\sum_{\substack{1\leq m\leq d}}x_mr_m+\\
&\sum_{\substack{ 1\leq i\leq j\leq d \\ id+j-\frac{i(i-1)}{2}\leq l}}
x_{id+j-\frac{i(i-1)}{2}} r_ir_j +r_{d+1}\mod q^N
\end{split}
\]
The MAC function consists of three types of terms. 
For a message  symbol $x_m$ with index $m$, one of the three types, as defined below, is
calculated. The final MAC is the sum of all the calculated terms.
\begin{enumerate}
\item  $x_m r_m$,  for  $1\leq m\leq d$;
\item  $x_mr_ir_j$,    for $d+1\leq m\leq l$ where  $m=id+j-\frac{i(i-1)}{2},$   and $1\leq i\leq j\leq d$;
\item  $r_{d+1}$, which is independent of message symbols. 
\end{enumerate}
For $d+1\leq m\leq l$,  the algorithm works as follows.\\
1. Consider the message symbols $m_{d+1},  m_{d+2}, \cdots m_l$ as a sequence;\\ 
2. Construct a key sequence using the product of a pair of key symbols $r_i$ and $r_j$ as follows:
start from the smallest $i=1, j=1$;  increase $j$ by one from $i$ to $d$; then increase $i$ by one and repeat to reach the highest values of
the two indexes.\\
3. Find the product of $x_m$ and the element of the key sequence constructed above, that corresponds with position $m$.

It can be seen that for a given pair $i$ and $j$, $m$ will satisfy $m=id+j-\frac{i(i-1)}{2}$.

\begin{lemma}
The probability that a computationally unlimited adversary can forge a message $({\bf x}', tag')$ with ${\bf x}'\neq {\bf x}$, that passes the verification test is no more than $\frac{2}{q^N}$.
\end{lemma}
We omit the security proof because of space and that it is essentially the same as the
proof of  Construction II.
\vspace{2mm}

\subsubsection{MAC Construction II} \label{MAC II}
We introduce a  MAC that can be seen as a different representation of Construction I above,
that  will  be used   in the construction of efficient randomized LV adversary code. 
The MAC can be described  by a set of equations over $F_q$. 
The {\it source state} of the MAC is a vector of length $Nl$ over $F_q$, 
\[
{\bf x}=\begin{bmatrix} 
x_{1,0},\cdots, x_{1, N-1}, \cdots, x_{l,0},\cdots, x_{l,N-1}
\end{bmatrix}^T
\] 
The {\it key} for the MAC is a vector of length $Nd+3N-2$ over $F_q$ where $d$ is the smallest integer satisfies $\frac{d(d+3)}{2}\geq l$,
\[
\begin{split}
{\bf r}=[ & r_{1,0},\cdots, r_{1,N-1}, r_{d,0}\cdots, r_{d,N-1}, \\
& r_{d+1,0},\cdots, r_{d+1,3N-3}]^T
\end{split}
\]
We write the key 
 in the form of an $(3N-2)\times (Nl+1)$ matrix:
\[
{\bf R}=\begin{bmatrix}
{\bf R}_1 \mid \cdots \mid & {\bf R}_d \mid & {\bf R}_{d+1} \mid  \cdots \mid & {\bf R}_{l} \mid & {\bf R}_{l+1} 
\end{bmatrix}
\] 
where ${\bf R}_m$ is a matrix that, depending on the value of the index $m$, can take the following forms.
For $1\leq m\leq d$, 
\[
{\bf R}_m=\begin{bmatrix}
r_{m,0} &0  &\cdots  &0 \\
r_{m,1} &r_{m,0}  &\cdots  &0 \\
\vdots &\vdots & \ddots & \vdots \\
r_{m,N-1} & r_{m,N-2} &\cdots  & r_{m,0}\\
0 &r_{m,N-1} &\cdots & r_{m,1} \\
\vdots &\vdots & \ddots & \vdots \\
0 &0 &\cdots  &r_{m,N-1}  \\
0 &0 &\cdots & 0\\
\vdots &\vdots & \ddots & \vdots \\
0 &0 &\cdots & 0\\
\end{bmatrix} 
\]
For $d+1\leq m\leq l$,
\[
{\bf R}_{m}=\begin{bmatrix}
r_{i,j,0} &0  &\cdots  &0 \\
r_{i,j,1} &r_{i,j,0}  &\cdots  &0 \\
\vdots &\vdots & \ddots & \vdots \\
r_{i,j,N-1} & r_{i,j,N-2} &\cdots  & r_{i,j,0}\\
r_{i,j,N} &r_{i,j,N-1} &\cdots & r_{i,j,1} \\
\vdots &\vdots & \ddots & \vdots \\
r_{i,j,2N-1} &r_{i,j,2N-2} &\cdots  &r_{i,N-1}  \\
0 &r_{i,j,2N-1} &\cdots & r_{i,j,N}\\
\vdots &\vdots & \ddots & \vdots \\
0 &0 &\cdots & r_{i,j,2N-1}\\
\end{bmatrix} 
\]
where $m$ is written as a pair of integers $i$ and $j$, similar to the description
of Construction I, and we have $r_{i,j,k}=\sum_ {\substack{ 0\leq a_1, a_2 \\  a_1+ a_2=k}}   
r_{i, a_1}r_{j, a_2}$ for $0\leq k\leq 2N-1$.

Finally, ${\bf R}_{l+1}=\left[
r_{d+1,0}, \cdots , r_{d+1,3N-3}
\right]^T
$.

The {\it tag} for a source state is a vector of length $3N-2$, 
\[
{\bf t}=[t_{0},\cdots, t_{3N-3}]^T.
\]

A source state $\bf x$ is encoded to the message $(\bf x, t)$ using the MAC algorithm,

\begin{equation}\label{maceq}
\begin{split}
&MAC({\bf x}, {\bf r})=\sum_{1\leq m\leq d}x_j{\bf R}_j+\sum_{d+1\leq m\leq l}x_{m}{\bf R}_{m} +{\bf R}_{l+1}\\
&=\left[
{\bf R}_1 \mid \cdots \mid  {\bf R}_{l} \mid  {\bf R}_{l+1} 
\right]\times\begin{bmatrix}
x_{1,0}\\
\vdots\\
x_{1, N-1}\\
\vdots\\
x_{l+1,0}\\
\vdots\\
x_{l+1,3N-3}\\
1
\end{bmatrix}
=\begin{bmatrix}
\bf t
\end{bmatrix}
\end{split}
\end{equation}

The verification algorithm  $Ver({\bf r},(\bf x', t'))$ for a key ${\bf r}$ is by calculating $MAC({\bf x'}, {\bf r})$,
and comparing it with the received ${\bf t'}$.

\begin{lemma}\label{lemma_MAC}
The probability that a computationally unlimited adversary can forge a message $(\bf x', t')$ with ${\bf x'\neq x}$, 
that passes  the verification is no more than $\frac{2}{q^{N}}$.
\end{lemma}
\begin{IEEEproof}
Appendix \ref{pf_MAC}.
\end{IEEEproof}

\section{Construction of LV Adversary Code}
In this section we describe the construction of an LV adversary code that uses the MAC algorithm in Section \ref{MAC II}
together with an FRS code with appropriately chosen parameters.

\subsection{$(N, q^{NuR}, \delta)$ randomized limited view adversary code}\label{SMT2}

{\em We assume the adversary reads $\rho N$ positions and adds errors to the same positions. }
Let $N$ and $R$ denote the code length and information rate, respectively.

The LV adversary code is over $F_q^u$. The sender $\cal S$ wishes to send the message ${\bf m}=(m_0,\cdots, m_{NuR-1}), m_i\in F_q$,  to the receiver.
\\

\noindent{\bf Randomized LV adversary code:}

{\setlength{\unitlength}{2mm}
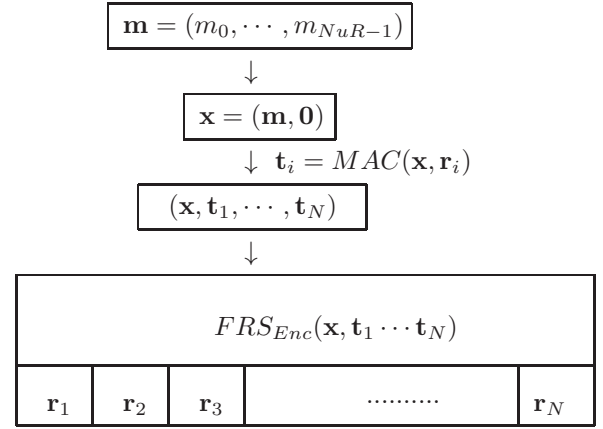
\begin{picture}(10,30)
\linethickness{0.2mm}
\put(8,25){\line(2,0){20}}
\put(8,25){\line(0,1){3}}
\put(8,28){\line(2,0){20}}
\put(28,25){\line(0,1){3}}
\put(9,26){${\bf m}=(m_0,\cdots, m_{NuR-1})$}
\put(17,23){$\downarrow$}
\put(13,19){\line(2,0){10}}
\put(13,19){\line(0,1){3}}
\put(13,22){\line(2,0){10}}
\put(23,19){\line(0,1){3}}
\put(14,20){${\bf x}=({\bf m},{\bf 0})$}
\put(17,17){$\downarrow$}
\put(19,17){${\bf t}_i=MAC({\bf x}, {\bf r}_i)$}
\put(10,13){\line(2,0){15}}
\put(10,13){\line(0,1){3}}
\put(10,16){\line(2,0){15}}
\put(25,13){\line(0,1){3}}
\put(12,14){$({\bf x},{\bf t}_1,\cdots, {\bf t}_N)$}
\put(17,11){$\downarrow$}
\put(2,0){\line(2,0){38}}
\put(2,0){\line(0,1){10}}
\put(2,4){\line(2,0){38}}
\put(2,10){\line(2,0){38}}
\put(40,0){\line(0,1){10}}
\put(7,0){\line(0,1){4}}
\put(12,0){\line(0,1){4}}
\put(17,0){\line(0,1){4}}
\put(35,0){\line(0,1){4}}
\put(4,1){${\bf r}_1$}
\put(9,1){${\bf r}_2$}
\put(14,1){${\bf r}_3$}
\put(25,1.5){$..........$}
\put(36,1){${\bf r}_N$}
\put(15,6){${FRS_{Enc}}({\bf x}, {\bf t}_1 \cdots {\bf t}_N)$}
\end{picture}

The LV adversary code is constructed over $F_q^u$ where $u=u_1+u_2$. 
The FRS code is over $F_q^{u_1}$ and the randomness ${\bf r}_i$ has length ${u_2}$. 
We set the parameters of MAC Construction II to be $l=\lceil uR \rceil$ and $d=\lceil \sqrt{2u_1} \rceil$.
 We have $u_2=Nd+3N-2=N\lceil\sqrt{2u_1}\rceil+3N-2$ and $u=u_1+N\lceil\sqrt{2u_1}\rceil+3N-2$.
\\

\noindent\textbf{Encoding algorithm performed by the sender $\cal S$ :}
\begin{framed}

\noindent{\bf Step 1:}  Append vector $\{{\bf 0}\} \in F_q^{N(l -uR)}$ to message ${\bf m}=(m_0,\cdots, m_{NuR-1})$, and form the vector ${\bf x}=\{{\bf m}, {\bf 0}\}$ of  length $Nl$. 
\\

\noindent{\bf Step 2:}  Generate random keys ${\bf r}_i, $$1\leq i\leq N$,  for the MAC Construction II.
Each key is written as a $(3N-2)\times (Nl+1)$ matrix,  
\[
\begin{split}
{\bf R}_i=[&{\bf R}_{i,1} \mid \cdots \mid  {\bf R}_{i, l} \mid {\bf R}_{i,d+1} ]
\end{split}
\] 

\noindent{\bf Step 3:} 
Use MAC Construction II to generate tags ${\bf t}_i=MAC({\bf x}, {\bf R}_i),$ $i=1,\cdots, N$ . 

\vspace{2mm}
\noindent
The FRS code is of dimension  $k=Nl+N(3N-2)$.
The message block for the FRS code is,
\[
\begin{split}
{\bf m}^{FRS}=({\bf x}, {\bf t}_1 \cdots {\bf t}_N)
\end{split}
\]

\noindent{\bf Step 4:}   Use the FRS encoding algorithm to encode $ {\bf m}^{FRS}$ to the codeword $c^{FRS}=Enc_{FRS}({\bf m}^{FRS})$.
 \\The  $i^{th}$ component of $c$, the  codeword of  the limited view adversary code, is 
obtained  by appending the  randomness ${\bf r}_i$ to  $c^{FRS}_i$, the $i^{th}$ component of the FRS code.
\[
c_i=(c^{FRS}_i, {\bf r}_i)
\] 
\end{framed}

\noindent\textbf{Decoding algorithm performed by the receiver $\cal R$ :}
\begin{framed}

\noindent{\bf Step 1:} Receive a corrupted word $y$ with the $i^{th}$ component $y_i=(y^{FRS}_i, \hat{{\bf r}}_i)$. 
Here $y^{FRS}_i$ and $\hat{{\bf r}}_i$ are the  $i^{th}$ component of the FRS code and the randomness in corrupted form, respectively.
\\

\noindent{\bf Step 2:} Use the FRS decoding algorithm to decode the FRS codeword $y^{FRS}$ and obtain the system of linear equations, \ref{eq_FRS2}.
\\

\noindent{\bf Step 3:} Generate $N$ systems of linear equations, each system obtained from the
set of linear equations generated from the
FRS decoding algorithm and one MAC key ${\bf r}_i$. The $i^{th}$ system of linear equation is of the form,

\begin{equation} \label{eq2}
\begin{split}
&\begin{bmatrix} 
{\bf B}_0 & {\bf B}_{1} &  \cdots & {\bf B}_{i} & \cdots & {\bf B}_{N} \\
{\bf R}'_i & {\bf 0} & \cdots&  -\bf I & \cdots & {\bf 0} \\
\end{bmatrix}\times\begin{bmatrix}
{\bf x}\\
{\bf t}_1\\
\vdots\\
{\bf t}_i\\
\vdots\\
{\bf t}_{N}\\
\end{bmatrix}=\begin{bmatrix}
-{\bf a}'\\
-{\bf R}_{i,d+1}\\
\end{bmatrix}
\end{split}
\end{equation}

The first $Nl+N(3N-2)$ equations are generated by the FRS decoding algorithm of Eq. \ref{eq_FRS2}:
the first $Nl$ columns of the matrix  of coefficients of these equations form ${\bf B}_{0}$,
 and for $1\leq i\leq N$,  columns $(Nl+(i-1)(3N-2))$ to $(Nl+i(3N-2)-1)$ of this matrix specify ${\bf B}_i$.
 Finally,   $-{\bf a}'$ is the right hand side  vector of Eq. \ref{eq_FRS2}. 
The last $3N-2$ equations are from MAC Construction II using key ${\bf r}_i$, with ${\bf R}'_i=[{\bf R}_{i,1} \mid \cdots \mid  {\bf R}_{i, l}]$, and $\bf I$ is identity matrix.  
\\

\noindent{\bf Step 4:}
Solves each of the $N$ systems of linear equations.
 Let ${\bf x}_i$ denote, the first $Nl$ components of a solution output by the $i^{th}$ system of linear equation. 
The $i^{th}$ system of linear equation is considered to have output output ${\bf x}_i$, if  ${\bf x}_i$  is the unique output 
of this system. 
Otherwise $\cal R$ marks the output of the $i^{th}$ system, as 
NULL.
 If there is a unique ${\bf x}$ output by a set  of the $N-\rho N$ systems 
 of linear equations, $\cal R$ outputs the first  $NuR$ components of that  ${\bf x}$  as ${\bf m}$.
\\ Otherwise outputs $\perp$.  

\end{framed}

\subsection{Adversary's reading and writing capability}

\begin{theorem} The $(N, q^{RN}, \delta)$ randomized limited view adversary code over $F_q^{u}$ above,  can correctly decode if the adversary reads and writes on the same set of size  $\rho N $  of a codeword. 
\[
\begin{split}
\rho \leq \min (&\frac{1}{2}-\frac{1}{2N},  \frac{v}{v+1}-\\
&\frac{v}{v+1}\frac{uR+3N}{N^2+u-N(\sqrt{N^2+2u}+3)-v} )
\end{split}
\]
\end{theorem}

\begin{IEEEproof}
Firstly, $\rho <1/2$: 
 If the adversary can read and write on half  of the components of a codeword $c$, 
 they can choose any other codeword $c'$ and add appropriate error vector to replace 
 components  of $c$  
 on the controlled positions to obtain $y$ which is equal to $c'$ on the controlled components,
and equal to $c$ on the remaining ones. The decoder can not decode $y$ and fail.

Secondly, we find a bound on $\rho$ when $\rho<\frac{1}{2}$.
The code dimension for the FRS code is $k=NuR$, and each component is in $F_q^u$.
Note that  only the FRS code, which is over $F_q^{u_1}$, contains the message information. Hence,  $k=Nu_1R_1$.
Let $R_{FRS}$ be the information rate of the FRS code. 
The decoding algorithm of LV adversary code need to satisfy the decoding condition of FRS code.  
According to Lemma \ref{le_fd}, the FRS code with length $N$ and information rate $R_{FRS}$ can decode $\rho N$ adversary errors if satisfying the condition:
\begin{eqnarray}\label{eq11}
       N-\rho N\geq N(\frac{1}{v+1}+\frac{v}{v+1}\frac{u_1R_{FRS}}{u_1-v+1})  
\end{eqnarray}
The   equation is satisfied if,
\[
N-\rho N\geq \frac{N}{v+1}+\frac{v}{v+1}\frac{(N(u_1R_1+1)+N(3N-2))}{u_1-v+1}
\]
The maximum error that the adversary can add is,
\[
\rho \leq \frac{v}{v+1}-\frac{v}{v+1}\frac{(u_1R_1+3N-1)}{u_1-v+1}
\]

The  LV adversary code is over $F_q^u$ and $u= u_1+\lceil\sqrt{2u_1}\rceil N+3N-2$. So we have,
\[
u_1\geq N^2+u-3N+1-N\sqrt{N^2+2u-2(3N-1)}
\]
The decoding condition of FRS code is satisfied if the following inequality is met:
\[
\begin{split}
\rho \leq &\frac{v}{v+1}-\frac{v}{v+1}\times\\
&\frac{uR+3N-1}{N^2+u-3N+2-N\sqrt{N^2+2u-2(3N-1)}-v+1}
\end{split}
\]
This is equivalent to,
\[
\rho \leq \frac{v}{v+1}-\frac{v}{v+1}\frac{uR+3N}{N^2+u-N(\sqrt{N^2+2u}-3)-v}
\]
\end{IEEEproof}

\subsection{Decoding error}

The adversary reads $\rho N$ components of a corrupted codeword and adds errors to the same positions using the 
knowledge of the components that are read. 

\begin{lemma}\label{theorem2}
If the adversary does not choose the  $i^{th}$ position for read and write,
the probability that the $i^{th}$ system of linear equations (Eqs. \ref{eq2}) does not produce 
the unique solution which contains the correct message ${\bf m}$ is at most 
 $\frac{2}{q^{N-v+1}}$. This is equivalent to,
\[
\begin{split}
\Pr[ &d_H({c'}^{FRS}, y^{FRS})\leq \rho N,  {\bf t}'_i=MAC({\bf x}', {\bf r}_i) |  {\cal C}[c_{i_1}\cdots c_{i_{\rho N}}]]\\
&\leq \frac{2}{q^{N-v+1}}
\end{split}
\]
with ${c'}^{FRS}=Enc_{FRS}({{\bf m}'}^{FRS})$ and ${{\bf m}'}^{FRS}=({\bf x}', {\bf t}'_1 \cdots {\bf t}'_N)$ and $({\bf x}'\neq {\bf x})$.
\end{lemma}
\begin{IEEEproof} 
Firstly, because the correct message is always contained in the decoded list of the FRS decoding algorithm, the correct ${\bf x}=\{{\bf m},{\bf 0}\}$ will be in the solution space 
of the system of linear Eq. \ref{eq2}. Also because the key ${\bf r}_i$ has not been modified, the solution will be  contained in the solution space of the equations generated by the MAC. 
Hence the solution space of the Eqs. \ref{eq2} must contain the correct message $\bf m$. 

Secondly,  a solution ${\bf x}' $,  where ${\bf x}'\neq {\bf x}$, of the system of linear Eqs. \ref{eq_FRS2} 
resulting from the FRS decoding algorithm, 
with probability 
at most $\frac{2}{q^N}$ will be a solution of the system of linear Eqs. \ref{eq2}. 
Now assume ${\bf x}'\neq {\bf x}$ is a solution of Eqs. \ref{eq2}.
This  means that it must satisfy the equations generated by MAC:
\begin{equation}\label{eq_MAC2}
\begin{bmatrix} 
{\bf R}'_i\   -\bf I 
\end{bmatrix}\\
\times\begin{bmatrix}
{\bf x}'\\
{\bf t}'_i\\
\end{bmatrix}=\begin{bmatrix}
-{\bf R}_{i,v+2}\\
\end{bmatrix}
\end{equation}
Using lemma \ref{lemma_MAC}, the probability that $MAC({\bf x}', {\bf r}_i)={\bf t}'_i$ is at most $\frac{2}{q^N}$.

Finally, 
the system of linear equations Eq. \ref{eq_FRS2} generated by the decoding algorithm  of the FRS code
produces a list of  at most $q^{v-1}$ solutions, $\{{c'}^{FRS}: d_H({c'}^{FRS}, y^{FRS})\leq \rho N\}$, where each codeword
represents a message of the form ${{\bf m}'}^{FRS}=({\bf x}', {\bf t}'_1 \cdots {\bf t}'_N)$.
The first $Nl$ components of each solution gives one solution for ${\bf x}'$. By union the probability of the solutions ${\bf x}'\neq {\bf x}$ of Eqs. \ref{eq_FRS2} that are also the solution of Eqs. \ref{eq_MAC2}, the Eqs. \ref{eq2} has more than one solution with probability no more than $\frac{2q^{v-1}}{q^N}$.

The adversary has no information of ${\bf r}_i$. After observing $\{c_{i_1},\cdots, c_{i_{\rho n}}\}$, the probabilty that there exist $\{{c'}^{FRS}: d_H({c'}^{FRS}, y^{FRS})\leq \rho N\}$ and the message passing MAC verification $MAC({\bf x}', {\bf r}_i)={\bf t}'_i$ is still equal to $\frac{2}{q^{N-v+1}}$.

\end{IEEEproof}

\begin{theorem}\label{theorem1}
The  decoding error of the $(N, q^{RN},\delta)$ randomized limited view adversary code is at most $\delta\leq \frac{2N}{q^{N-v+1}}$. 
\end{theorem}
\begin{IEEEproof}
Let $y=Enc({\bf m},r)+e$ be  the corrupted word,  and $I_3 = S_r=S_w$ denote the positions that are read and modified by the 
adversary. For a codeword $c'=(c'^{FRS}, {\bf r}'_1,\cdots, {\bf r}'_N)$ with ${c'}^{FRS}=Enc_{FRS}({{\bf m}'}^{FRS})$ and ${{\bf m}'}^{FRS}=({\bf x}', {\bf t}'_1 \cdots {\bf t}'_N)$ and ${\bf x}'\neq {\bf x}$, let $I^{c'}_1=\{i : c'_i=y_i\}$ and $I^{c'}_2=\{i : MAC({\bf x}', {\bf r}'_i)={\bf t}'_i\}$.

According to definition \ref{def_generaldecodingerror2}, the probability of decoding failure for an encoding of a message ${\bf m}$ that satisfies the observation set
$(c_{i_1}\cdots c_{i_{\rho N}})$ is,

\[
\begin{split}
&\Pr[
B(Enc({\bf m}, r)+e, \rho N)
\cap \{C\setminus{C}^{{\bf m}}\}  \neq \emptyset  |\   {\cal C}[c_{i_1}\cdots c_{i_{\rho N}}]]\\
\end{split}
\]

This is the probability that for a codeword  $c'\in C\setminus{C}^{{\bf m}}$, there exists 
two subsets $I_1^{c'}$ and  $I_2^{c'}$ such that,  $|I^{c'}_1|\geq N-\rho N$, $|I^{c'}_2|=N$ and $  |I^{c'}_1\cap I^{c'}_2|\geq N-\rho N $.
The latter two conditions imply $|I^{c'}_1\cap I^{c'}_2|\geq \rho N +1$ if $\rho< \frac{1}{2}$, which can be written as, $ |\{[N]\setminus I_3\}\cap I^{c'}_1\cap I^{c'}_2|=1 $.

Note that $|I^{c'}_1|\geq N-\rho N$  implies  $d_H({c'}^{FRS}, {y}^{FRS})\leq\rho N$, and $ |\{[N]\setminus I_3\}\cap I^{c'}_1\cap I^{c'}_2|=1 $ implies existence of $i^{c'}$ such that $i^{c'} \in \{ I^{c'}_1\cap I^{c'}_2\}$ and  $i^{c'}  \in [N]\setminus I_3$.

This means that we have,
\[
\begin{split}
&\Pr[
B(Enc({\bf m}, r)+e, \rho N)
\cap \{C\setminus{C}^{{\bf m}}\}  \neq \emptyset  |\   {\cal C}[c_{i_1}\cdots c_{i_{\rho N}}]]\\
\leq &  \Pr[ (i^{c'} \in [N]\setminus I_3), (i^{c'} \in \{ I^{c'}_1\cap I^{c'}_2\}) , \\
&  (d_H({c'}^{FRS}, {y}^{FRS})\leq\rho N)  \;|\;\ {\cal C}[c_{i_1}\cdots c_{i_{\rho N}}] ]\\
\leq &  (N-\rho N)\Pr[ (i^{c'}\notin I_3), (i^{c'} \in \{ I^{c'}_1\cap I^{c'}_2\}) , \\
&  (d_H({c'}^{FRS}, {y}^{FRS})\leq\rho N)  \;|\;\ {\cal C}[c_{i_1}\cdots c_{i_{\rho N}}] ]\\
= &  (N-\rho N)\Pr[ (i^{c'}\notin I_3),  (MAC({\bf x}',{\bf r}_i)={\bf t}'_i), \\
& (d_H({c'}^{FRS}, {y}^{FRS})\leq\rho N) \;|\;\ {\cal C}[c_{i_1}\cdots c_{i_{\rho N}}] ]\\
\leq & \frac{2N}{q^{N-v+1}}
\end{split}
\]
The last inequality is correct because of lemma \ref{theorem2}.

\end{IEEEproof}

If we choose $v=\frac{1}{\varepsilon}$, $u=\frac{2}{\varepsilon^4}+\frac{2N}{\varepsilon^2}$  where $\varepsilon>0$ is a small value,  the decoding capability $\rho$ can be approximated is $\rho=\min(\frac{1}{2}-\frac{1}{2N}, 1-(1+N\varepsilon^2)R-N\varepsilon^4 -N^2\varepsilon^6)$,  and the decoding error will be given by $\delta \leq q^{\frac{1}{\varepsilon}-N}$.
The field size $q$ can be chosen as  the smallest prime $q>Nu$. The encoding algorithm is polynomial in $N$. For decoding algorithm, the computational complexity of solving any $i^{th}$ system of linear equation Eqs. \ref{eq2} is $\mathcal{O}(((uN+N^2)\log q)^2)$ and there are $N$ systems of linear equations. So the  computational time of decoding algorithm is  polynomial in $\mathcal{O}(N((uN+N^2)\log q)^2)$.

\begin{corollary}
Assume the adversary is allowed to read  (at most) $\rho $ fraction of a codeword and can write on the same set. The $(N, q^{RN},\delta)$ randomized LV adversary code over $F_q^{\frac{2}{\varepsilon^4}+\frac{2N}{\varepsilon^2}}$ with,
\[
\rho \leq \min\left(\frac{1}{2}-\frac{1}{2N}, 1-(1+N\varepsilon^2)R-N\varepsilon^4 -N^2\varepsilon^6\right)
\]
can  correctly decode the errors 
and the decoding error  $\delta \rightarrow 0$ if $N\rightarrow \infty$. The computational time is polynomial in $N$.

\end{corollary}

The construction above can be immediately used to construct an optimal 1-round $\delta$-RMT, by using the encoding algorithm of the LV adversary code with appropriate length, to 
construct  a codeword for the message,
 and simply send the $i^{th}$ component of the codeword on path $i$ in the RMT setting. The decoding error in LV adversary codes is equivalent to the strongest definition of reliability in RMT scenario where the adversary can choose the message, and so $\delta$ in RMT will be at most equal to the decoder failure in LV adversary codes.
The optimality follows from the constant (non-zero) rate of the LV adversary code.
\begin{corollary}
The construction of the randomized LV adversary code give an optimal  1-round $\delta$-RMT, where $\delta$ is the same as the decoding error in LV adversary codes.
\end{corollary}

\section{Concluding Remarks}

We introduced randomized limited view adversary codes and gave an efficient construction  that wiith appropriate choice of parameters, can correct close to $N/2$ errors and will have information rate close to $1/2$.
Although in general the observation and corruption sets can be different, in our construction we assumed they are the same. Giving a construction without this assumption will be our future work. 
In our construction the field size is a function of $N$ and so small $\delta$ can be obtained for large field sizes. Finding good LV adversary codes with fix field size, and/or information rate approaching $1-\rho-\varepsilon$ are open problems.

Randomized codes do not have the restrictions of deterministic codes on their parameters and
can achieve much better performance (higher $\rho_r$ and $\rho_w$ for fixed $R$).  
  Finding general bounds and relationship
among the information rate $R$, observation $\rho_r$ and corruption $\rho_w$ ratios, and
finding the information capacity of LV adversary codes remain important research questions.

Our work showed that LV adversary codes provide a more refined way of modelling RMT scenarios allowing to 
cater for the information rate of these protocols.
Extending definition of LV adversary codes to interactive scenarios will be an interesting open question.

\begin{appendix}

\subsection{Decoding algorithm of FRS code} \label{decode_FRS}

Linear algebraic list decoding \cite{Gur11} has two main steps: interpolation and message finding as outlined below.

\begin{itemize}
\item Find a polynomial, $Q(X, Y_1, \cdots, Y_v)=A_0(X)+A_1(X)Y_1+\cdots+A_v(X)Y_v$, over $F_q$ such that $\mbox{deg}(A_i(X)) \leq D$, for $i=1 \cdots v$, and $\mbox{deg}(A_0(X)) \leq D+k-1$, 
  satisfying $Q(\alpha_i, y_{i_1}, y_{i_2},\cdots ,y_{i_v})=0$ for $1\leq i\leq n_0$, where $n_0=(u_1-v+1)N$.

\item Find all polynomials $f(X) \in F_q[X]$ of degree at most $k-1$, with  coefficients $f_0, f_1 \cdots f_{k-1}$,
    that satisfy, $A_0(X)+A_1(X)f(X)+A_2(X)f(\gamma X)+\cdots+A_v(X)f(\gamma^{v-1}X)=0$,
by solving linear equation system.
\end{itemize}

The two above requirements are satisfied if $f \in F_q[X]$ is a polynomial of degree at most $k- 1$ whose FRS encoding (Eq \ref{FRS_encode}) agrees with the received word $\bf y$ in at least $T$ components: 
\[
T>N(\frac{1}{v+1}+\frac{v}{v+1}\frac{u_1R}{u_1-v+1})
\]

This means we need to find all polynomials $f(X) \in F_q[X]$ of degree at most $k-1$, with coefficients $f_0, f_1, \cdots, f_{k-1}$, that satisfy, 
\[
\begin{split}
&A_0(X)+A_1(X)f(X)+A_2(X)f(\gamma X)+\cdots+\\
&A_v(X)f(\gamma^{v-1}X)=0
\end{split}
\]

Let us denote $A_i(X) = \sum_{j=0}^{D+k-1} a_{i,j}X^j$
for $0 \leq i \leq v$. ($a_{i,j} = 0$ when $i \geq 1$ and $j \geq D$). 
Define the polynomials,
\[
\begin{cases}
\begin{split}
&B_0(X)  =  a_{1,0} + a_{2,0}X + a_{3,0}X^2+ \cdots + a_{v,0}X^{v-1}\\
& \ \ \ \ \ \ \vdots\\
&B_{k-1}(X) = a_{1,k-1} + a_{2,k-1}X + a_{3,k-1}X^2+ \cdots + \\
&\ \ \ \ \ \ \ \  \ \ \ \ \ \ \ \ a_{v,k-1}X^{v-1}\\
\end{split}
\end{cases}
\]

We examine the condition that the coefficients of $X^i$ of the polynomial $Q(X) = A_0(X) + A_1(X)f(X) + A_2(X)f(\gamma X) + \cdots + A_v(X)f(\gamma^{v-1} X)=0$ equals $0$, for $i=0\cdots k-1$. This is equivalent to the following system of linear equations for $f_0\cdots f_{k-1}$.

\begin{equation}\label{eq_FRS2}
\begin{split}
&\begin{bmatrix}
B_0(\gamma^0) &0  & 0 &\cdots  &0 \\
B_1(\gamma^0) &B_0(\gamma^1)  & 0 &\cdots  &0 \\
B_2(\gamma^0) &B_1(\gamma^1)  & B_0(\gamma^2) &\cdots  &0 \\
\vdots &\vdots  & \vdots & \vdots & \vdots \\
B_{k-1}(\gamma^0) & B_{k-2}(\gamma^{1}) &B_{k-3}(\gamma^2)  &\cdots  & B_0(\gamma^{k-1})
\end{bmatrix}\\
&\times\begin{bmatrix}
f_0\\
f_1\\
f_2\\
\vdots\\
f_{k-1}
\end{bmatrix}=\begin{bmatrix}
-a_{0,0}\\
-a_{0,1}\\
-a_{0,2}\\
\vdots\\
-a_{0,k-1}
\end{bmatrix}
\end{split}
\end{equation}
The rank of the matrix of Eqs. \ref{eq_FRS2} is at least $k-v+1$ because there are at most $v-1$ solutions of equation $B_0(X)=0$ so at most $v-1$ of $\gamma^i$ that makes $B_0(\gamma^i)=0$. The dimension of solution space is at most $v-1$ because the rank of matrix of Eqs. \ref{eq_FRS2} is at least $k-v+1$. So there are at most $q^{v-1}$ solutions to Eqs. \ref{eq_FRS2} and this determines the size of the list which is equal to $q^{v-1}$.

\subsection{Proof of lemma \ref{lemma_MAC}}\label{pf_MAC}

\begin{IEEEproof}
We need to find the following probability:
$$\Pr[ (MAC({\bf x}', {\bf r})={\bf t}') | (MAC({\bf x}, {\bf r})={\bf t})]$$

The MAC function given by Eqs. \ref{maceq}, is equivalent to the MAC of the polynomial form in Eq.  \ref{eq_MAC}. For  $0\leq i\leq 3N-3$, the  coefficients of $X^i$ in both sides of equation \ref{eq_MAC} form the same 
equation as the $i^{th}$ equation in the system of linear equations \ref{maceq}.  
\begin{equation}\label{eq_MAC}
\begin{split}
& t(X)=MAC({\bf x}, {\bf r})=\sum_{1\leq m\leq d}x_m(X)r_m(X)+\\
&\sum_{\substack{d+1\leq m\leq l\\ m=id+j-\frac{i(i-1)}{2}}}x_{m}(X) r_i(X)r_j(X) +r_{d+1}(X)\mod q
\end{split}
\end{equation}
where each polynomial is given below
\begin{eqnarray*}
\begin{split}
&x_m(X)=x_{m,0} + \cdots +x_{m,N-1} X^{N-1}\mod q,\,\, 1\leq i\leq l\\
&r_m(X)=r_{m,0}+\cdots +r_{m,N-1}X^{N-1}\mod q,\,\, 1\leq m\leq d\\
&r_{m}(X)=r_{i,j,0}+\cdots +r_{i,j,2N-2}X^{2N-2}=\\
&r_i(X)r_j(X)\mod q,\,\, d+1\leq m\leq l,\,\, m=id+j-\frac{i(i-1)}{2}\\
&r_{d+1}(X)=r_{d+1,0}+\cdots+ r_{d+1,3N-3}X^{3N-3}\mod q
\end{split}
\end{eqnarray*}
Finally, $t(X)= t_{0}+\cdots+ t_{3N-3}X^{3N-3}\mod q$. 

So if we can prove that the adversary's forging capability to the MAC in the form of Eq. \ref{eq_MAC} is no more than $\varepsilon$, then the the adversary's forging capability to MAC construction II (Eqs. \ref{maceq}) is also no more than $\varepsilon$.

Next we prove the adversary forging capability to MAC in the form of Eq. \ref{eq_MAC} is no more than $\frac{2}{q^N}$. Assume the adversary forges a message $({\bf x}', {\bf t}')$ with ${\bf x}'\neq {\bf x}$,  that passes the verification. 
We  write the MAC in polynomial form.
\begin{equation}
\begin{split}
& t'(X)=MAC({\bf x}', {\bf r})=\sum_{1\leq m\leq d}x'_m(X)r_m(X)+\\
&\sum_{\substack{d+1\leq m\leq l\\ m=id+j-\frac{i(i-1)}{2}}}x'_{m}(X) r_i(X)r_j(X) +r_{d+1}(X)\mod q
\end{split}
\end{equation}
By subtracting the two equations we will have,
\[
\begin{split}
&\sum_{\substack{d+1\leq m\leq l\\ m=id+j-\frac{i(i-1)}{2}}}\Delta x_{m}(X) r_i(X)r_j(X)+ \\
&\sum_{1\leq m\leq d}\Delta x_m(X)r_m(X)=\Delta t(X) \mod q
\end{split}
\]
The above equation has at most $2q^{N(d-1)}$ solutions for $(r_1(X),\cdots, r_d(X))$. This means that there are at most $2q^{N(d-1)}$ keys $\bf r$ that satisfy $MAC({\bf x}, {\bf r})={\bf t}$, and $MAC({\bf x}', {\bf r})={\bf t}'$. However, there are $q^{Nd}$ possible values for $\bf r$ satisfying  $MAC({\bf x}, {\bf r})={\bf t}$. So the success probability of the forgery is, 
\[
\begin{split}
&\Pr[  (MAC({\bf x}', {\bf r})={\bf t}') | (MAC({\bf x}, {\bf r})={\bf t})]\\
=&\frac{2q^{N(d-1)}}{q^{Nd}}=\frac{2}{q^N}
\end{split}
\] 
\end{IEEEproof}

\end{appendix}

\end{document}